\documentclass[aps,12pt,final,onecolumn,superscriptaddress]{revtex4}%
\usepackage{amsfonts}
\usepackage{amsmath}
\usepackage{amssymb}
\usepackage{graphicx}%
\providecommand{\U}[1]{\protect\rule{.1in}{.1in}}

\begin{document}
\title[Effect of superthermality on nonlinear electrostatic modes in plasmas]{Effect of superthermality on nonlinear electrostatic modes\\in plasmas}
\author{S.~Sultana}
\email{ssultana02@qub.ac.uk}
\affiliation{Centre for Plasma Physics, Queen's University Belfast, BT7 1NN Northern
Ireland, UK}
\author{A.~Danehkar}
\email{adanehkar01@qub.ac.uk}
\affiliation{Centre for Plasma Physics, Queen's University Belfast, BT7 1NN Northern
Ireland, UK}
\author{N.\,S.~Saini}
\email{ns.saini@qub.ac.uk}
\affiliation{Centre for Plasma Physics, Queen's University Belfast, BT7 1NN Northern
Ireland, UK}
\author{M.\,A.~Hellberg}
\email{hellberg@ukzn.ac.za} \affiliation{School of Physics,
University KwaZulu-Natal, Durban 4000, South Africa}
\author{I.~Kourakis}
\email{i.kourakis@qub.ac.uk}
\affiliation{Centre for Plasma Physics, Queen's University Belfast, BT7 1NN Northern
Ireland, UK}

\begin{abstract}
The nonlinear propagation of electron-acoustic solitary structures is
investigated in a plasma containing $\kappa$ distributed (superthermal)
electrons. Different types of localized structures are shown to exist. The
occurrence of modulational instability is investigated.

\end{abstract}
\maketitle

\section{Introduction \label{introduction}}

Superthermal particles in laboratory, space and astrophysical
plasmas are often modeled by a $\kappa$-type distribution function
(df) \cite{Vasyliunas1968,Sultana2010}. The superthermality
parameter $\kappa$ measures the deviation from a Maxwellian
distribution (the latter is recovered for infinite $\kappa$). Our
twofold aim here is to investigate the effect of superthermality on
electrostatic solitary waves, and also on self-modulated
wavepackets.

Electron-acoustic waves (EAW) occur in plasmas containing two
distinct temperature electron populations (here referred to as
``cold'' and ``hot'' electrons) \cite{Watanabe1977,Mace1990}. These
are high frequency electrostatic electron oscillations, where the
restoring force comes from the hot electrons pressure and the cold
electrons provide the inertia \cite{Watanabe1977,Verheest2007},
while ions plainly provide a neutralizing background. The phase
speed $v_{ph}$ of the EAW is much larger than the thermal speeds of
both cold electrons and ions but much smaller than the cold
electrons (i.e., $v_{ph,c}$, $v_{ph,i}\ll v_{ph} \ll v_{ph,h}$).
EAWs survive Landau damping in the region $T_{h}/T_{c} \geq10$ and
$0.25 \leq n_{c0}/n_{h0} \leq4$ \cite{Watanabe1977,Mace1990}, where
we have defined the temperature ($T_{c}$, $T_{h}$) and density
($n_{c}$, $n_{h}$) of the electron constituents (`c' for cold, `h'
for hot).

\section{Electron fluid model \label{electronfluidmodel}}

We consider a three component plasma consisting of inertial (\textquotedblleft
cool\textquotedblright) electrons, $\kappa$-distributed (\textquotedblleft
hot\textquotedblright) electrons and stationary background ions. In a 1D
geometry, the dynamics of the cold electrons is governed by the following
normalized equations:%
\[%
\begin{array}
[c]{cc}%
\dfrac{\partial n}{\partial t}+\dfrac{\partial(nu)}{\partial x}=0, & \text{
\ \ }\dfrac{\partial u}{\partial t}+u\dfrac{\partial u}{\partial x}%
=\dfrac{\partial\phi}{\partial x}-\dfrac{\sigma}{n}\dfrac{\partial P}{\partial
x},
\end{array}
\]%
\[%
\begin{array}
[c]{cc}%
\dfrac{\partial P}{\partial t}+u\dfrac{\partial P}{\partial x}+3P\dfrac
{\partial u}{\partial x}=0, & \text{ \ \ }\dfrac{\partial^{2}\phi}{\partial
x^{2}}=-(\beta+1)+n+\beta\left(  1-\dfrac{\phi}{\kappa-\tfrac{3}{2}}\right)
^{-\kappa+1/2},
\end{array}
\]
We have scaled all relevant physical quantities as: cold electron density
$n=n_{c}/n_{c0}$; fluid speed $u=u_{c}/v_{0}$; electric potential $\phi
=\Phi/\Phi_{0}$; time $t=t\omega_{pc}$; space $x=x/\lambda_{0}$; pressure
$P=P/n_{c0}k_{B}T_{c}$; we have defined: $v_{0}=(k_{B}T_{h}/m_{e})^{1/2}$,
$\lambda_{0}=(k_{B}T_{h}/4\pi n_{c0}e^{2})^{1/2}$ and $\omega_{pc}^{-1}=(4\pi
n_{c0}e^{2}/m_{e})^{-1/2}$, and also the density- and temperature- ratio(s):
$\beta=n_{h,0}/n_{c,0}$ and $\sigma=T_{c}/T_{h}$.%

\begin{figure}
[ptb]
\begin{center}
\includegraphics[
height=2.1534in,
width=5.3705in
]%
{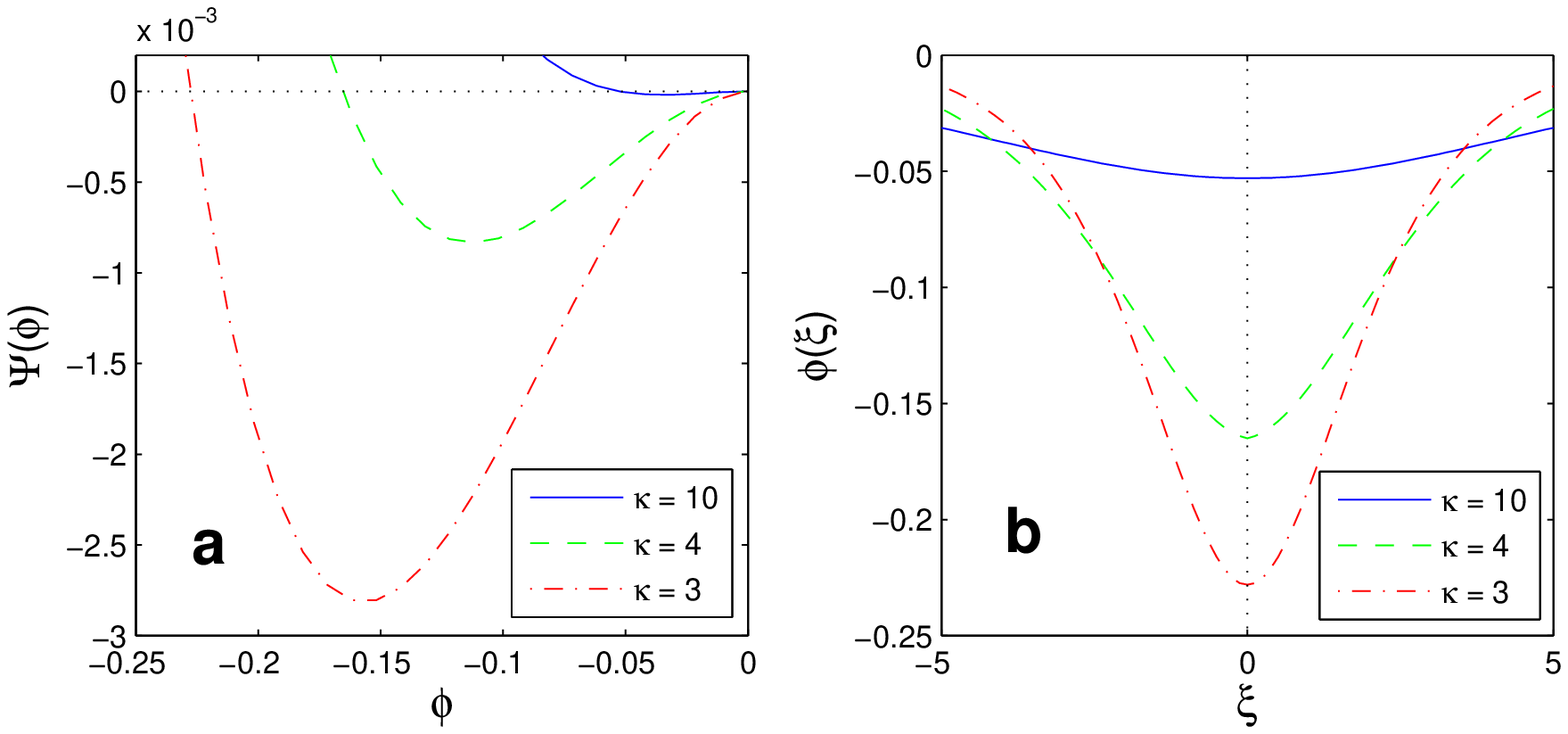}%
\caption{Variation of the pseudopotential $\Psi(\phi)$ with $\phi$
(left); the electric potential $\phi$ vs. $\xi$ (right). We have
considered various values
of k, and $\sigma=0.01$, $\beta=1$, and $M=1$.}%
\label{fig1}%
\end{center}
\end{figure}

\section{Arbitrary amplitude solitary excitations}

Anticipating stationary profile localized excitations, we shift from
variables $\{x,t\}$ to $x=x-Mt$, where $M$ is the solitary wave
speed, scaled by $v_{0}$
(defined above). We obtain $u=M\left(  1-\dfrac{1}{n}\right)  $, $u={M-(M}%
^{2}{+2\phi-3n^{2}\sigma+3\sigma)}^{1/2}$, and $P=n^{3}$. Poisson's equation
thus leads to a pseudo-energy balance equation
\begin{equation}
\frac{1}{2}\left(  \frac{d\phi}{d\xi}\right)  ^{2}+\Psi(\phi)=0,
\label{eq2_37}%
\end{equation}
where the ``Sagdeev'' pseudopotential function $\Psi(\phi)$ reads
\cite{Danehkar2011}
\begin{align}
\Psi(\phi)=  &  (1+\beta)\phi+\beta\left[  1-\left(  1+\frac{\phi}%
{-\kappa+\tfrac{3}{2}}\right)  ^{-\kappa+3/2}\right]  +\frac{1}{6\sqrt
{3{\sigma}}}\left[  \left(  {M+}\sqrt{3{\sigma}}\right)  ^{3}-{{\left(
{M-}\sqrt{3{\sigma}}\right)  ^{3}}}\right. \nonumber\\
&  -\left(  {2\phi+}\left[  {M+}\sqrt{3{\sigma}}\right]  ^{2}\right)
^{3/2}\left.  +{\left(  {2\phi+\left[  {M-}\sqrt{3{\sigma}}\right]  ^{2}%
}\right)  }^{3/2}\right]  \,. \label{eq2_38}%
\end{align}

\subsection{Soliton Existence}

\begin{figure}
[ptb]
\begin{center}
\includegraphics[
width=2.8in
]%
{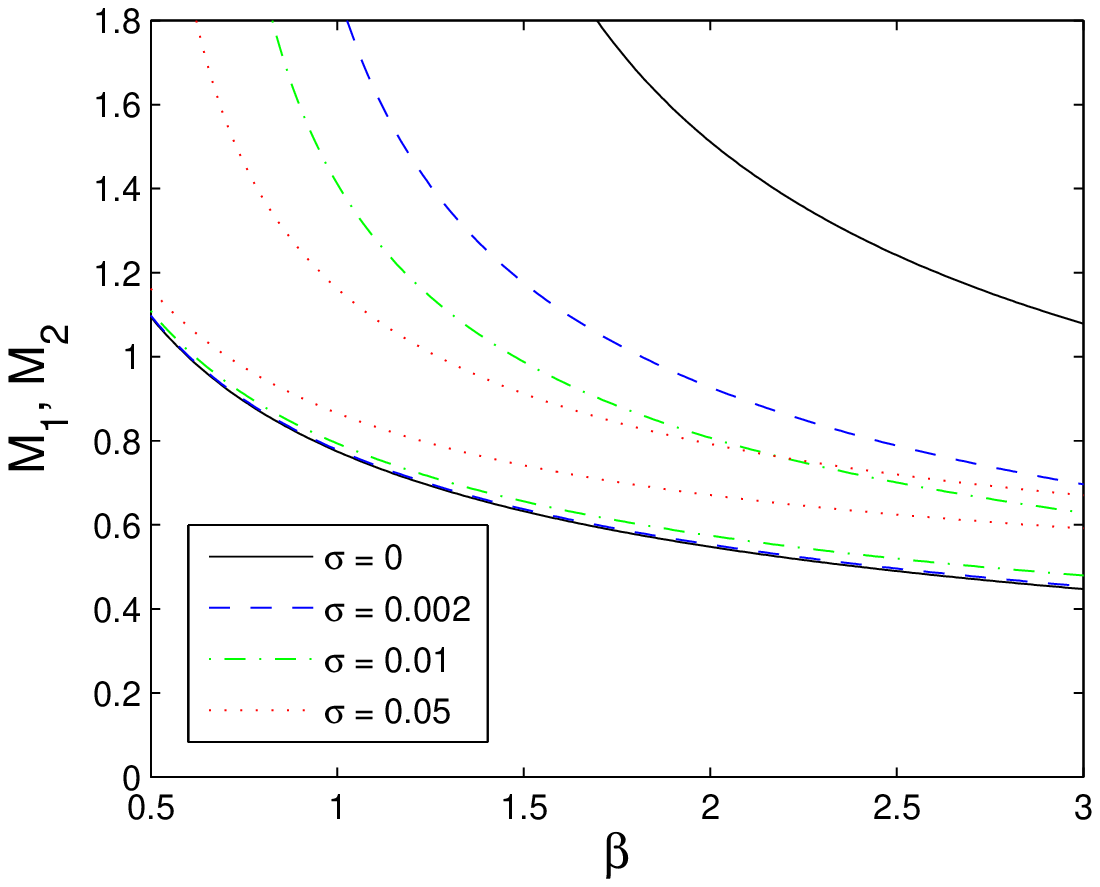}%
\ \ \ \ \
\includegraphics[
width=2.8in
]%
{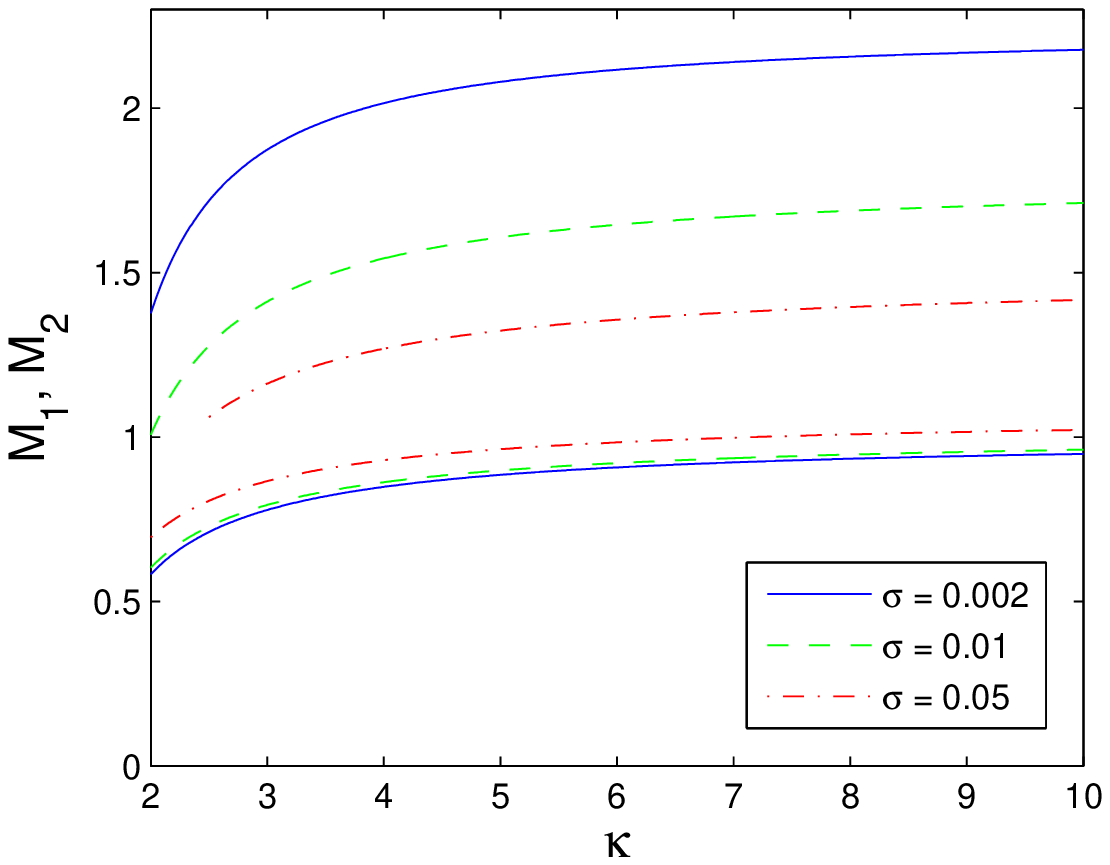}%
\caption{Soliton existence region ($M_{1}<M<M_{2}$) for different
temperature
ratio $\sigma$ values, versus $\beta$ for $\kappa=3$ (left panel);
versus $\kappa$ for $\beta=1$ (right panel).}%
\label{fig2}%
\end{center}
\end{figure}

In order for solitons to exist, we need to impose \cite{Verheest2007}:
$\Psi^{\prime}(\phi=0)=0$ and $\Psi^{\prime\prime}(\phi=0)<0$ (where the prime
denotes differentiation wrt $\phi$), leading to the (true sound speed)
threshold $M_{1}=\left[  \frac{\kappa-3/2}{\beta(\kappa-1/2)}+3\sigma\right]
^{1/2}$. An upper limit for $M$ is obtained by imposing the reality
requirement \cite{Danehkar2011}. $F_{2}(M)=\Psi(\phi)|_{\phi=\phi_{\max}}>0$
(where $\phi_{\max}$ is a limit on the electrostatic potential value;
$\Psi(\phi)$ is real for $\phi_{\max}<\phi$; see Fig. \ref{fig1}a). The region
thus obtained is depicted in Figure \ref{fig2}.%

\section{Modulated electron-acoustic wavepackets}

We consider small ($\varepsilon\ll1$) deviations of all state
variables, say $S$ ($=n,u,\phi$), from the equilibrium state, viz.
$S={S}^{(0)}+\Sigma
_{n=1}^{\infty}\varepsilon^{n}\,\sum_{l=-n}^{n}S^{(nl)}e^{il(kx-\omega
t)}$, and allow for a weak space-/time-dependence of the $l$--th
harmonic amplitudes $S^{(nl)}$. In what follows, we ignore the
pressure term (\textit{cold electron model}) for simplicity, and set
$\alpha={n_{c,0}}/{n_{h,0}}$ ($=\beta^{-1}$).
\begin{figure}
[ptb]
\begin{center}
\includegraphics[width=2.5in
]%
{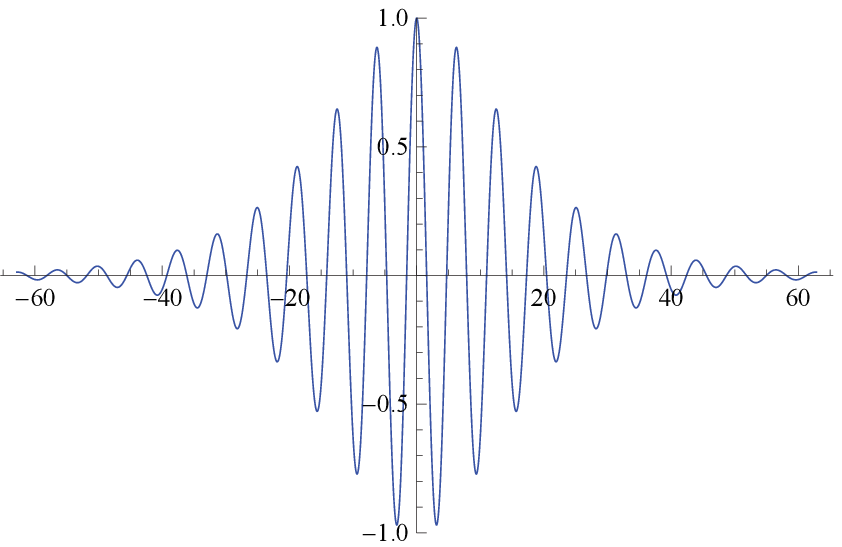}%
\ \ \ \ \
\includegraphics[width=2.5in
]%
{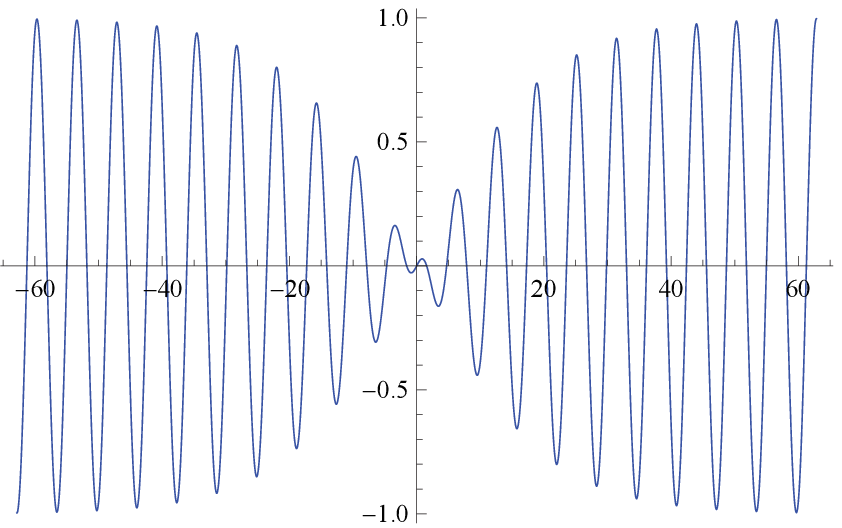}%
\caption{Envelope type solitary excitations:
bright type (left panel) and dark type (right panel).}%
\label{fig3}%
\end{center}
\end{figure}

The 1st order ($\sim\varepsilon^{1}$) expressions provide the EAW
\textit{dispersion relation}
$\omega^{2}=\frac{k^{2}\alpha}{k^{2}+c_{1}}$, along with the
amplitudes of the first harmonics. The 2nd and 0th harmonics are
obtained at order $\varepsilon^{2}$. Annihilation of secular terms
at 3rd order yields a nonlinear Schr\"{o}dinger (NLS) type equation:
\begin{equation}
i\, \frac{\partial\psi}{\partial\tau} + P \, \frac{\partial^{2} \psi}%
{\partial\zeta^{2}} + Q \, |\psi|^{2}\,\psi= 0 \, .\label{nlse}%
\end{equation}
where the amplitude $\psi\equiv\phi_{1}^{(1)}(\zeta, \tau)$ depends
on $\zeta= \varepsilon(x - v_{g} t)$, $\tau= \varepsilon^{2} t$,
while $v_{g}= \frac{d \omega }{d k} = \frac{\omega^{3}c_{1}}{k^{3}
\alpha}$ and $P$ and $Q$ are dispersion and nonlinear coefficients
respectively.
\begin{figure}
[ptb]
\begin{center}
\includegraphics[width=2.24in
]
{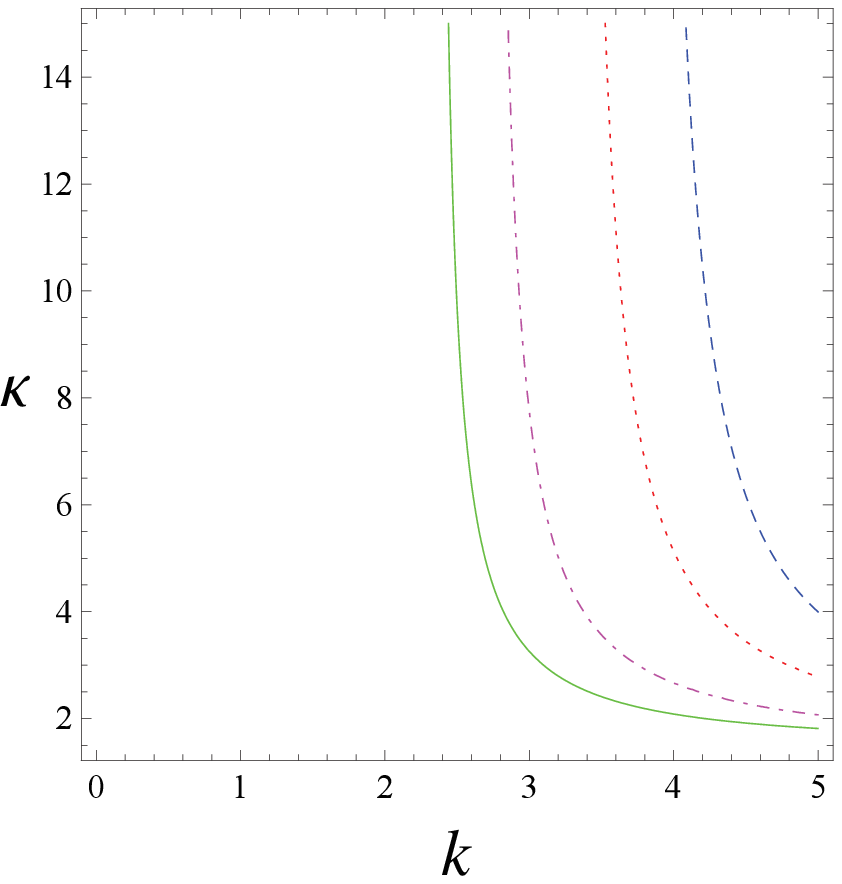}%
\ \ \ \ \
\includegraphics[width=2.3in
]%
{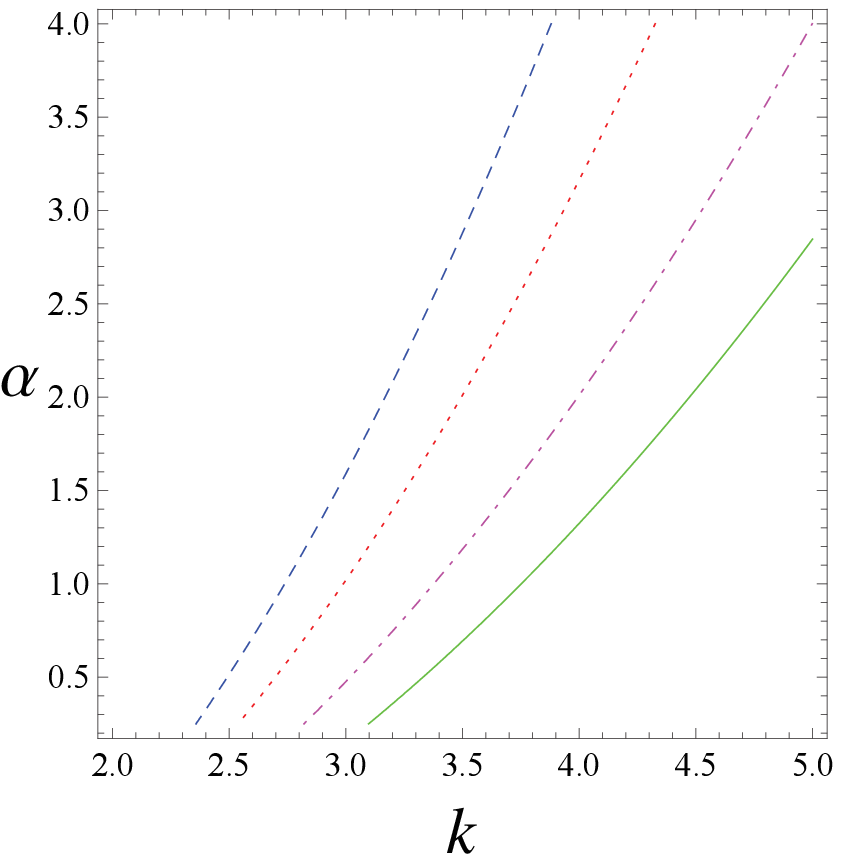}%
\caption{$PQ = 0$ (or $k = k_{cr}$) contours \textit{vs} carrier
wavenumber $k$ and superthermality parameter $\kappa$, or density
ratio $\alpha$. Left panel: $\alpha=0.25$ for the green curve; $1$
for magenta; $2.5$ for red; and $4$
for blue. Right panel: $\kappa=3$ for green; $4$ for magenta; $8$ for red; and $100$ for blue.}%
\label{fig4}%
\end{center}
\end{figure}
\begin{figure}
[ptb]
\begin{center}
\includegraphics[width=2.5in
]
{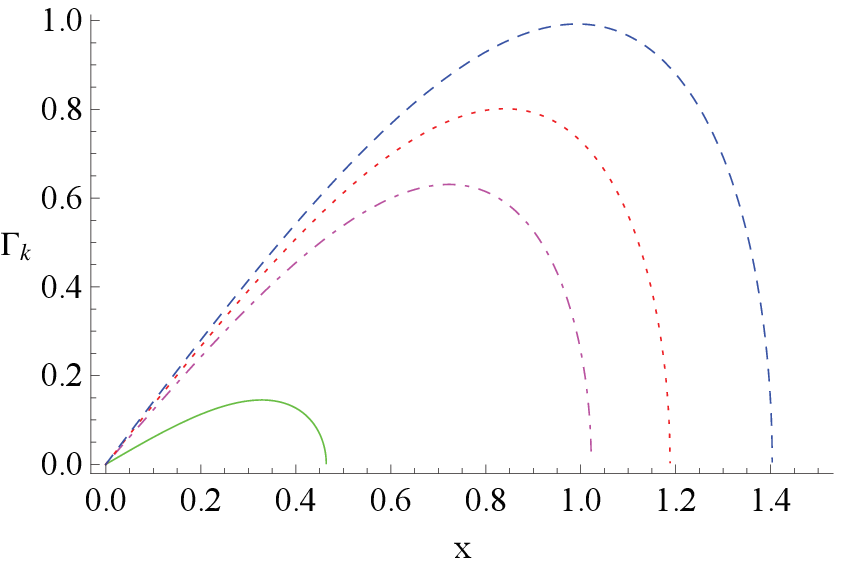}%
\ \ \ \ \
\includegraphics[width=2.5in
]%
{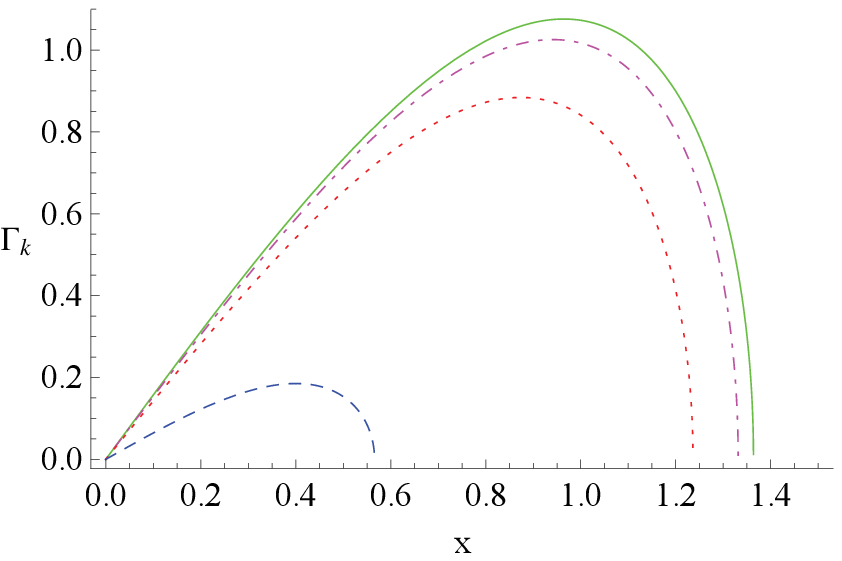}%
\caption{Modulational instability growth rate (normalized by its
value for infinite $\kappa$) versus perturbation wavenumber. Left
panel: $\kappa = 100$, $7$, $5$, $3.5$ (top to bottom) for $\alpha =
0.5$, $k = 3.2$. Right panel: $\alpha = 0.5$, $1$, $2$, $4$ (top to
bottom)
for $k = 4.5$, $\kappa = 7$.}%
\label{fig5}%
\end{center}
\end{figure}

\subsection{Modulational instability}

Adopting standard procedure \cite{Sultana2010}, we investigate the
occurrence of modulational instability by considering a harmonic
solution of (\ref{nlse}) and then a harmonic amplitude perturbation
with (wavenumber, frequency)=($\tilde{k},\tilde{\omega}$). A
nonlinear dispersion relation is thus obtained:
$\tilde{\omega}^{2}=P^{2}\tilde{k}^{2}(\tilde{k}^{2}-2\frac{Q}{P}|
\tilde {\psi}_{1,0}| ^{2})$. Provided that $PQ>0$, wavenumbers above
$\tilde{k}_{cr}=\left(  \frac{2Q}{P}\right) ^{1/2}| \tilde{\psi
}_{1,0}| $ lead to (amplitude) modulational instability (MI). For
$PQ<0$, wavepackets are stable.

\subsection{Envelope solitons}

In the modulationally stable region ($PQ < 0$, in fact essentially
for large wavelengths) ``\textit{dark}'' solitons may occur, i.e.
exact solutions in the form \cite{Kourakis2004}: $\psi = \psi_0 \{1
- d^2 {\rm sech}^2[(\zeta -V\tau)/L]\}$. On the other hand, for $PQ
> 0$, ``\textit{bright}'' envelope solitons occur in the form \cite{Kourakis2004}: $\psi =
\psi_0 {\rm sech}[(\zeta -V\tau)/L]$. In the above, $\psi_0$ is the
asymptotic electric potential amplitude value, $V$ is the
propagation speed and $L$ is the soliton width, while the positive
constant $d$ regulates the depth of the void ($d = 1$ for black
solitons or $d < 1$ for grey ones).

\section{Summary}

Stronger superthermality leads to higher amplitude solitary
excitations (as suggested by Fig. \ref{fig1}). Both the cold
electron temperature and concentration significantly effect on the
soliton existence domain, as the upper Mach number limit $M_{2}$
increases for higher ``cold'' electron temperature, while the sonic
threshold $M_{1}$ is decreased for higher $n_{c0}$ (see Fig.
\ref{fig4}). The modulational instability growth rate may be reduced
due to stronger superthermality (see Fig. \ref{fig5}a), while
(somewhat counter-intuitively) the presence of more excess hot
(superthermal) electrons increases the instability growth rate (see
Fig. \ref{fig5}b).

\section*{Acknowledgements}

This work was supported by a UK EPSRC Science and Innovation award
to Queen's University Belfast Centre for Plasma Physics (Grant No.
EP/D06337X/1). The work of AD was supported via postgraduate
scholarship at Queen's University Belfast from the Department for
Employment and Learning (DEL) Northern Ireland.

\end{document}